\documentclass[fl eqn,twoside]{article}
\date{}
\usepackage{amsmath}
\usepackage{amsfonts}
\usepackage{amssymb}
\usepackage{graphicx}
\def\be{\begin{equation}}
\def\ee{\end{equation}}
\def\be{\begin{appendix}}
\def\ee{\end{appendix}}

\begin{document}
\date{Today}
\title{{\bf{Generalized uncertainty principle in resonant detectors of gravitational waves }}}

\author{
{\bf {\normalsize Sukanta Bhattacharyya}$^{a}$,\thanks{sukanta706@gmail.com}}
{\bf {\normalsize Sunandan Gangopadhyay}
$^{b}$\thanks{sunandan.gangopadhyay@gmail.com,
sunandan.gangopadhyay@bose.res.in}},
{\bf {\normalsize Anirban Saha }
$^{a}$\thanks{anirban@associates.iucaa.in}}\\
$^{a}$ {\normalsize Department of Physics, West Bengal State University, Barasat, Kolkata 700126, India}\\
$^{b}$ {\normalsize  Department of Theoretical Sciences,}\\{\normalsize S.N. Bose National Centre for Basic Sciences,}\\{\normalsize JD Block, 
Sector III, Salt Lake, Kolkata 700106, India}\\[0.1cm]
}
\date{}
\maketitle\
\begin{abstract}
\noindent With the direct detection of gravitational waves by advanced LIGO detector, a new ``window" to quantum gravity phenomenology has been opened. At present, these detectors achieve the sensitivity to detect the length variation ($\delta L$), $\mathcal{O} \approx 10^{-17}-10^{-21}$ meter. Recently a more stringent upperbound on the dimensionless parameter $\beta_0$, bearing the effect of generalized uncertainty principle has been given which corresponds to the intermediate length scale $l_{im}= \sqrt{\beta_0} l_{pl} \sim 10^{-23} m$. Hence the flavour of the generalized uncertainty principle can be realised by observing the response of the vibrations of phonon modes in such resonant detectors in the near future. In this paper, therefore, we calculate the resonant frequencies and transition rates induced by the incoming gravitational waves on these detectors in the generalized uncertainty principle framework. It is observed that the effects of the generalized uncertainty principle bears its signature in both the time independent and dependent part of the gravitational wave-harmonic oscillator Hamiltonian. We also make an upper bound estimate of the GUP parameter.

\end{abstract}

\section{Introduction}
Quantum mechanics and general relativity are two revolutionary theories which give the present description of the fundamental laws of nature. It is also realised that for an understanding of the universe near the Planck epoch, a suitable quantum theory of gravity is needed.  
String theory \cite{ad,kk}, loop quantum gravity \cite{lqg1,lqg2}, noncommutative geometry \cite{gi}, are frameworks that try to put forward a quantum theory of gravity. In the last few decades there has been extensive studies in these areas to build up a theory of quantum gravity (QG), and all these studies imply the existence of an observer independent minimum length scale, namely, the Planck length $l_{pl} \approx 10^{-33} cm$. In fact the existence of a minimal length, namely, the Planck length can be argued from the requirement of modifying the Heisenberg uncertainty principle (HUP) to the generalized uncertainty principle (GUP).  The idea that 
if we do not disregard gravity, quantum theory does prevent the measurability of some field value in an arbitrarily small region was first proposed in 
\cite{bronstein, bronstein1}. The possible connection between gravitation and fundamental length was further elaborated in \cite{mead} where an analysis of the effect of gravitation on hypothetical experiments was carried out indicating that it is impossible to measure the position of a particle with error less than $\Delta x \geq \sqrt{G}= 1.6\times 10^{-33}$ cm, where $G$ is the Newton's gravitational
constant in natural units. This strong indication of the existence of the GUP from the different realms of QG theory and also from gedanken experiments motivates us to investigate different aspects in theoretical physics like black hole physics \cite{mmg}-\cite{sbbh}, quantum gravity corrections in various quantum systems such as particle in a box, Landau levels, simple harmonic oscillator \cite{sdas,sdas3}, path integral representation of a particle moving under a potential in the GUP framework \cite{sp,sb} and so on. Various thought experiments leads to the following simplest form of the GUP \cite{kempf} 
\begin{eqnarray}
\Delta q_{i}\Delta p_{i}~\geq~\frac{\hbar}{2}\left[1+\beta(\Delta p^{2}+\langle
p\rangle^{2})+2\beta(\Delta p_{i}^{2}+<p_{i}>^{2})\right];~~~~i=1,2,3 \label{deltaxi}
\label{GUP}
\end{eqnarray}
where $p^2= \Sigma^3_{j=1} p_j p_j$ and $q_j$, $p_j$ are the position and its conjugate momenta. The dimension of the  GUP parameter $\beta$ defined as $\beta=\frac{\beta_0}{(M_{p}c)^2}$ is $(momentum)^{-2}$, where $M_{p}=1.2\times 10^{19}GeV$ is the Planck mass. 
Studies involving harmonic oscillators with minimal length uncertainty relations have been carried out in \cite{id}-\cite{pb11}. It is also quite natural that the order of the dimensionless parameter $\beta_0$ would play a crucial role for realizing such effects of the GUP. Indeed a lot of effort has been put to find an upper bound of the GUP parameter $\beta_0$ \cite{fs1, sdas}, \cite{bawaj}-\cite{fs}. A numerical estimate of the GUP parameter was made in \cite{vgnas}.
Probing possible deviations from the quantum commutation relation due to Planck scale physics has also been done using quantum optics \cite{pikov}-\cite{kumar}.
In spite of these works, the fact remains that the testing of the GUP is extremely challenging and therefore initiates the proposal of a realistic experimental set up to test the GUP.

\noindent A new window in exploring high energy phenomena has been opened with the direct detection of gravitational waves (GWs) \cite{bbh, bns} by the LIGO \cite{lig} and VIRGO \cite{vir} detectors. The idea of observing GWs was sowed by J. Weber with the introduction of resonant bar detectors \cite{jw1,jw2} in 1960s, and a huge effort has been spent in increasing their sensitivity as much as possible. Present day bar detectors \cite{GW-detection_status}
are capable of detecting the fractional variations $\Delta L$ 
of the bar-length $L$ down to $\frac{\Delta L}{L} \sim 10^{-19}$, with $L\sim 1$ metre,
which may be sensitive enough to allow us to probe the effects of QG.
A lot of work has been carried out to probe the footprints of noncommutativity \cite{sb2}-\cite{sg5} in these detectors. The sub-millikelvin cooling
of the normal modes of the ton-scale AURIGA GW detector has been exploited to place an upper limit on the GUP parameter by looking at possible Planck-scale modifications on the ground-state energy of an oscillator \cite{fm, fma}. Potential tests of the GUP in the LIGO detector have also been considered in \cite{rbm}.
  
\noindent  Motivated by the above discussion, in this paper we present the quantum mechanical effects of the GW bar detectors in the GUP framework. In resonant detectors incoming GW interacts with elastic matter. This interaction causes tiny vibrations called phonons. The amplitudes of phonons are many order smaller than the nuclear size. These vibrations are described as the quantum mechanical forced harmonic oscillator (HO). Hence the physical description of these detectors is nothing but a quantum mechanical forced GW-HO system. Therefore, in this paper we construct the Hamiltonian of the above system using the GUP algebra (\ref{GUP}).  We treat the effects of GW and the GUP as perturbation on the HO-system and calculate the formal perturbative solutions of that system. First we apply the time independent perturbation bearing the effects of only the GUP and get the perturbed eigenstates of the $1$-dimensional HO with the new perturbed energy eigenvalues. After that we calculate the time dependent perturbation to calculate the transitions between the states of the HO caused by the GWs containing the GUP signature. The results show that the resonant frequencies get modified by the GUP parameter. The number of transitions and their amplitudes also reveal the presence of the GUP. Therefore, this mathematical construction of the GW-HO system in presence of the GUP shows that it can serve as a good candidate of realising the GUP effect in GW detection data.


The organization of the paper is as follows. In section 2, we briefly outline how the HO-GW interaction can be modelled in presence of the GUP and obtain the relevant Hamiltonian. The complete perturbative calculation to obtain the working formula for the transition probabilities among the shifted energy levels for a generic GW waveform is presented in section 3. In section 4 we use the various GW waveforms to calculate the corresponding transition probabilities and discuss the possibilities of detecting the GUP signature. We conclude in section 6. We also have an Appendix at the end which presents a brief derivation of the Newtonian force equation of the detector coordinates.


\section{The GUP GW-HO interaction model}
\noindent The GW interacts with elastic matter in the plane (considered here as the $x-y$ plane) perpendicular to the propagating vector of the GW (taken in $z$-direction). Hence the interaction of GW with matter can be described as a $2-$dimensional HO-GW interaction. Therefore, first we take the geodesic deviation equation for a $2-$dimensional HO of mass $m$ and intrinsic frequency $\varpi$ in a proper detector frame as 
\begin{equation}
m \ddot{{q}} ^{j}= - m{R^j}_{0,k0} {q}^{k} - m \varpi^{2} q^{j}~; ~j=1,2
\label{e5}
\end{equation}
in terms of the components of the curvature tensor 
${R^j}_{0,k0} = - \frac{d {\Gamma^j}_{0k}}{d t}  = -\ddot{h}_{jk}/2  $
\footnote{The dot denotes derivative with respect to the coordinate time of the proper detector frame. It is the same as its proper time to first order in the metric perturbation and ${q}^{j}$ is the proper distance of the pendulum from the origin.}. Here metric perturbation $h_{\mu \nu}$ takes the form 
\begin{eqnarray}
g_{\mu\nu} = \eta_{\mu\nu} + h_{\mu\nu}~; \, |h_{\mu\nu}|<<1
\label{metric_perturbation}
\end{eqnarray}
on the flat Minkowski background $\eta_{\mu\nu}$. Before proceeding further lets discuss about the validity and gauge conditions applied in eq.(\ref{e5}). This equation is valid only when the spatial velocities involved are non-relativistic 
and $|{q}^{j}|$ is much smaller than the reduced wavelength $\frac{\lambda}{2\pi}$ of the GW. These conditions are obeyed by the resonant bar detectors and earth bound interferometric detectors. It is to be noted that the transverse-traceless (TT) gauge conditions are applied to remove the unphysical degrees of freedom. Hence only two relevant components, namely the $\times$ and $+$ polarizations of the GW arises in the curvature tensor ${R^j}_{0,k0} = -\ddot{h}_{jk}/2$. This condition actually makes the spatial part of the GW to be unity ( $e^{i \vec{k}.\vec{x}} \approx 1$) all over the detector site in case of the plane-wave expansion of GW.
Thus, the GW interaction give rise to a time-dependent piece in eq.(\ref{e5}). Now $h_{jk}$ containing the polarization information reads
\begin{equation}
h_{jk} \left(t\right) = 2f \left(\varepsilon_{\times}\sigma^1_{jk} + \varepsilon_{+}\sigma^3_{jk}\right)
\label{e13}
\end{equation}
\noindent where $\sigma^1$ and $\sigma^3$ are the Pauli spin matrices, $2f$ is the amplitude of the GW and $\left( \varepsilon_{\times}, \varepsilon_{+} \right)$ are the two possible polarization states of the GW satisfying the condition $\varepsilon_{\times}^2+\varepsilon_{+}^2 = 1$ for all $t$. The frequency $\Omega$ is contained in the time dependent amplitude $2f(t)$  for linearly polarized GW, whereas the time dependent polarization states $\left( \varepsilon_{\times} \left(t \right), \varepsilon_{+} \left( t \right) \right)$ contains the frequency $\Omega$ for the circularly polarized GW.

\noindent Now the classical Lagrangian describing the geodesic eq.(\ref{e5}) upto a total derivative can be recast as\footnote{Note that since the coordinates of the detector are non-relativistic (as has been elaborated in the Appendix), covariant and contravariant notations are equivalent, that is, $q_i =q^i$.}
\begin{equation}
{\cal L} = \frac{1}{2} m\dot {q_{j}}^2 - m{\Gamma^j}_{0k}
\dot {q}_{j} {q}^{k}  - \frac{1}{2} m \varpi^{2}  q_{j}^2~
\label{e8}
\end{equation}
where $q_{j}^2= q_{j}q_{j}$ with summation implied on $j$. 

\noindent Using the canonical momentum ${p}_{j} = m\dot {q}_{j} - m {\Gamma^j}_{0k} {q}^{k}$ corresponding to ${q}_{j}$, the Hamiltonian of the GW-HO system reads
\begin{equation}
{H} = \frac{1}{2m}\left({p}_{j} + m {\Gamma^j}_{0k} {q}^{k}\right)^2 + \frac{1}{2} m \varpi^{2} q_{j}^2  .
\label{e9}
\end{equation}
\noindent With this background in place, to probe the GUP we have to do the quantum mechanical description of the same with the GUP modified Heisenberg algebra. Therefore we follow the standard prescription of quantum mechanics by lifting the phase-space variables  $\left( q^{j}, p_{j} \right)$ to operators  $\left( {\hat q}^{j}, {\hat p}_{j} \right)$ in the GUP framework. Now the inequality (\ref{GUP}) is equivalent to the following modified Heisenberg algebra \cite{kempf}
\begin{eqnarray}
[\hat q_{i}, \hat p_{j}]=i\hbar(\delta_{ij}+\beta\delta_{ij}\hat{p}^{2}+2\beta \hat{p}_{i}\hat{p}_{j})~.
\label{GUP2}
\end{eqnarray}
Note that the Jacobi identity  $[ \hat q_{i},\hat q_{j}]=0=[\hat p_{i}, \hat p_{j}]$ ensures the above commutation relation. Next we can define the position and momentum operators upto first order in $\beta$ obeying eq.(\ref{GUP2}) as 
\begin{eqnarray}
\hat{q}_{i}={q}_{0i}~~~~~,~~~~~\hat{p}_{i}={p}_{0i}(1+\beta {p}_{0}^{2}) 
\label{GUPR}
\end{eqnarray}
where $q_{0i},~p_{0j}$ satisfy the usual canonical commutation relations $[q_{0i}, p_{0j}]=i\hbar\delta_{ij}$. Using this map, the Hamiltonian (\ref{e9}) describing the GW-HO system in presence of the GUP  up to first order in $\beta$ can be recast as 
\begin{eqnarray}
H=\frac{p_{0j}^2}{2m}+\frac{\beta}{m}p_{0j}^2p_{0n}^2+\frac{1}{2} \Gamma^j_{0l}\left(p_{0j}q_{0}^l+q_{0}^lp_{oj}\right)+\frac{\beta}{2}\Gamma^j_{0l}\left(p_{0j}p_{on}p_{0n}q_{0}^l+q_{0}^lp_{0j}p_{on}p_{on}\right)+\frac{1}{2} m \omega^2 q_{oj}^2 +\mathcal O(\beta^2) ~.
\label{3}
\end{eqnarray}
In the subsequent discussion, we shall consider the resonant bar detectors as a 
one-dimensional\footnote{A typical bar is a cylinder of length $L\equiv 3$ m and radius $R \equiv 30$ cm, so in a first approximation we can treat its vibrations 
as one-dimensional.} system \cite{Magg}. Hence we analyze the dynamics of a one-dimensional HO in presence of the GUP interacting with the GW. For notational simplicity we use $\hat p_{0j} \equiv p$ and $\hat q_{0j}\equiv q$. Therefore, the Hamiltonian describing the same in one-dimension up to first order in $\beta $ reads 
\begin{eqnarray}
H=\left(\frac{p^2}{2m}+\frac{1}{2} m \omega^2 q^2 \right)+\frac{\beta}{m}p^4+\frac{1}{2} \Gamma^1_{01}\left(pq+qp\right)+\frac{\beta}{2}\Gamma^1_{01}\left(p^3q+qp^3\right)~.
\label{4}
\end{eqnarray}
Now we can break the Hamiltonian (\ref{4}) as
\begin{eqnarray}
H=H_{0}+H_{1}+H_{2}
\label{5b}
\end{eqnarray}
where
\begin{eqnarray}
H_0&=&\frac{p^2}{2m}+\frac{1}{2}m\omega^2q^2 \nonumber\\
H_1&=&\frac{\beta}{m}p^4  \nonumber\\
H_2&=&\frac{1}{2} \Gamma^1_{01}\left(pq+qp\right)+\frac{\beta}{2}\Gamma^1_{01}\left(p^3q+qp^3\right)~.
\label{5a}
\end{eqnarray}
\noindent Here $H_0$ stands for the Hamiltonian of ordinary HO while $H_{1}$ and $H_{2}$ are the time independent \cite{sdas} and time dependent part of the Hamiltonian respectively. It is to be noted that $H_1$ and $H_2$ are small compared to $H_0$. Here $H_{1}$ arises from the kinetic part of the particle due to the presence of the GUP. Thus, $H_{1}$ does not contain explicit time dependence and according to quantum mechanics time independent perturbation makes shift in the energy eigenvalues with new perturbed eigenstates.
On the other hand the first bracketed term in $H_{2}$ shows the pure GW effect and the second one contains the effect of both GUP and GW. Now  transition between the states of the HO occurs due to the time dependent part of the perturbed Hamiltonian. In this paper we calculate the transition rates due to $H_{2}$, containing the effects of both the GUP and GW between the perturbed states.


\noindent To do this we define the momentum and the position operators in terms of the raising and lowering operators as
\begin{eqnarray}
p&=&-i\left(\frac{\hbar m \omega}{2}\right)^\frac{1}{2}\left(a-a^\dagger\right)\nonumber\\
q&=&\left(\frac{\hbar}{2 m \omega}\right)^\frac{1}{2}\left(a+a^\dagger\right). 
\label{6}
\end{eqnarray}
Using these and recalling the relation ${\Gamma^i}_{0j}=\frac{1}{2}\dot h_{ij}$ which in turn implies ${\Gamma^1}_{01}=\frac{1}{2}\dot h_{11}$, 
the Hamiltonian in eq.(\ref{5a}) can be recast as
\begin{eqnarray}
H_0&=&\hbar \omega\left(a^\dagger a +\frac{1}{2}\right) \nonumber
\end{eqnarray}
\begin{eqnarray}
H_{1}=\frac{\beta}{m}\left(\frac{\hbar m \omega}{2}\right)^2\left[aaaa-aaaa^\dagger-aaa^\dagger a+aaa^\dagger a^\dagger-aa^\dagger aa+aa^\dagger a a^\dagger+aa^\dagger a^\dagger a-a a^\dagger a^\dagger a^\dagger  \right. \nonumber\\  \left.
-a^\dagger aaa+a^\dagger aa a^\dagger +a^\dagger a a^\dagger a -a^\dagger a a^\dagger a^\dagger +a^\dagger a^\dagger a a-a^\dagger a^\dagger a a^\dagger -a^\dagger a^\dagger a^\dagger a+a^\dagger a^\dagger a^\dagger a^\dagger\right] \nonumber
\end{eqnarray}
\begin{eqnarray} 
H_{2}= i \hbar \dot h_{11}\left[-\frac{1}{2}\left(aa-a^\dagger a^\dagger\right)+\frac{\beta \hbar m \omega}{4}\left(aaaa-aa a^\dagger a-aa^\dagger aa+a a^\dagger a^\dagger a
-a^\dagger aa a^\dagger+a^\dagger a a^\dagger a^\dagger +a^\dagger a^\dagger a a^\dagger  \right. \right.
\nonumber\\ \left. \left. -a^\dagger a^\dagger a^\dagger a^\dagger\right)\right] ~.
\label{14}
\end{eqnarray}
We shall make use of these relations in the next section to obtain the perturbative energy levels and then find the transition probabilities between them.


\section{Perturbed energy levels and transitions}
In this section we proceed to calculate the perturbed eigenstates due to time independent Hamiltonian $H_1$. Using time independent perturbation theory, the perturbed eigenstates read
\begin{eqnarray}
|0 \rangle^\beta &=& |0 \rangle + \frac{\Delta}{8}~~\left[6 \sqrt{2}~|2 \rangle -\sqrt{6}~|4 \rangle \right] \nonumber\\
|2 \rangle^\beta &=& |2 \rangle + \frac{\Delta}{8}~~\left[-6\sqrt{2}~|0 \rangle +28 \sqrt{3}~|4 \rangle -3 \sqrt{10}~|6 \rangle \right] \nonumber\\
| 4\rangle^\beta &=& |4 \rangle + \frac{\Delta}{8}~~\left[\sqrt{6}~|0 \rangle - 28 \sqrt{3}~|2 \rangle +22 \sqrt{30}~|6 \rangle - 2 \sqrt{105}~|8 \rangle \right]
\label{8}
\end{eqnarray}
with the corresponding energies
\begin{eqnarray}
E_0^\beta&=&\left(\frac{1}{2}+\frac{3}{4} \Delta \right) \hbar \omega \nonumber\\
E_2^\beta&=& \left(\frac{5}{2}+\frac{39}{4} \Delta \right) \hbar \omega \nonumber\\
E_4^\beta &=& \left(\frac{9}{2}+\frac{123}{4} \Delta \right) \hbar \omega~.
\label{9}
\end{eqnarray}
Here $\Delta= \beta \hbar m \omega$ is the dimensionless parameter showing the GUP effect.

\noindent With the modified states due to the GUP in place, we are now ready to investigate the transitions between the various states of the HO due to the incoming GWs. 

\noindent Now to the lowest order of approximation in time dependent perturbation theory, the probability amplitude of transition from an initial state $|i\rangle$ to a final state $|f \rangle$, ($i\neq f$), due to a perturbation {\bf{$\hat{H}_2(t)$}} is given by \cite{kurt}
\begin{eqnarray}
C_{i \rightarrow f}(t\rightarrow \infty)  =  -\frac{i}{\hbar} \int_{-\infty}^{t\rightarrow +\infty} dt' F \left( t' \right) e^{\frac{i}{\hbar}(E_f -E_i)t'} \langle \Psi_f | \hat{Q}|\Psi_i \rangle
\label{probamp}
\end{eqnarray}
where $\hat{H}_2(t)=F(t)\hat{Q}$ with $F(t)=\dot h_{11} $, and $\hat{Q}$ is given by
\begin{eqnarray}
\hat{Q}=i \hbar \left[-\frac{1}{2}\left(aa-a^\dagger a^\dagger\right)+\frac{\beta \hbar m \omega}{4}\left(aaaa-aa a^\dagger a-aa^\dagger aa+a a^\dagger a^\dagger a
-a^\dagger aa a^\dagger+a^\dagger a a^\dagger a^\dagger +a^\dagger a^\dagger a a^\dagger  \right. \right.
\nonumber\\  \left. \left. -a^\dagger a^\dagger a^\dagger a^\dagger\right)\right]  ~.
\label{9a}
\end{eqnarray} 
Note that for an ordinary HO only $|0\rangle\rightarrow |2\rangle$ transition will occur. But due to the presence of the GUP, we get another transition 
$|0\rangle^\beta\rightarrow |4\rangle^\beta$ in addition to the previous one $(  |0\rangle^\beta \rightarrow |2\rangle^\beta  )$ with different amplitudes. 

\noindent In this work, two transitions namely; $0^\beta\rightarrow 2^\beta$ and $0^\beta\rightarrow 4^\beta$ have been observed. Therefore, using eq.(\ref{9a}) 
in eq.(\ref{probamp}), we find the transition amplitudes to be 
\begin{eqnarray}
C_{0^\beta \rightarrow 2^\beta } =A \int_{-\infty}^{t\rightarrow +\infty} dt'~~ \dot h_{11}~~e^{i(2+9 \Delta)\omega t'} \nonumber\\
C_{0^\beta \rightarrow 4^\beta} =B \int_{-\infty}^{t\rightarrow +\infty} dt'~~ \dot h_{11}~~e^{i(4+30 \Delta)\omega t'}
\label{10}
\end{eqnarray}
where $A= \left(\frac{1}{\sqrt{2}}+\frac{9}{4 \sqrt{2}} \Delta\right)$ and{\bf{ $B= -3 \sqrt{6}\Delta $ }}are dimensionless constants. In the limit $\beta \rightarrow 0$, we get the transitions for ordinary HO in $1$-dimension interacting with GWs. 

\noindent Eq.(\ref{10}) is one of the main findings in this paper. These transition amplitudes show that the presence of the GUP can be checked by measuring the corresponding transition probabilities from the relation
\begin{eqnarray}
P_{i\rightarrow f}=  |C_{i\rightarrow f}|^{2}.
\label{tp}
\end{eqnarray}
\noindent In the next section, we shall calculate the transition amplitudes for different types of incoming GWs.


\section{Transition probabilities for different types of GW templates}
We now look at different GW templates. In reality, the actual form of the GW signals are very complicated. Hence we start with some simple forms of GW templates generated from different astronomical events. 

\subsection{Periodic linearly polarized GW}
\noindent First we consider the simple type of periodic GW with linear polarization. This has the form
\begin{equation}
h_{jk} \left(t\right) = 2f_{0} \cos{\Omega t} \left(\varepsilon_{\times}\sigma^1_{jk} + \varepsilon_{+}\sigma^3_{jk}\right)
\label{lin_pol}
\end{equation}
where the amplitude varies sinusoidally with a single frequency $\Omega$. In this case, we get the transition probabilities to be
\begin{eqnarray}
P_{0^\beta\rightarrow 2^\beta} &=& \left(2 \pi f_0 \Omega A \epsilon_+\right)^2 \times \left[ \delta\left( \omega\left(2 +9 \Delta \right)+ \Omega \right) -\delta\left( \omega\left(2 +9 \Delta \right)- \Omega \right) \right]^2 \nonumber\\
P_{0^\beta\rightarrow 4^\beta} &=&  \left(2 \pi f_0 \Omega B \epsilon_+\right)^2 \times \left[\delta \left( \omega \left(4 +30 \Delta \right)+\Omega \right)- \delta \left( \omega \left(4 +30 \Delta \right)-\Omega \right) \right]^2~.
\label{trans_prob_lin_pol1}
\end{eqnarray}
The frequency $\omega$ of the resonant bar must lie in the physical range $\left(0<\omega< \infty \right)$. Hence eq.(\ref{trans_prob_lin_pol1}) takes the form
\begin{eqnarray}
P_{0^\beta\rightarrow 2^\beta} &=& \left(2 \pi f_0 \Omega A \epsilon_+\right)^2 \times \left[\delta\left( \omega \left(2 +9  \Delta \right)- \Omega \right)  \delta(0)\right] \nonumber\\
P_{0^\beta\rightarrow 4^\beta} &=&  \left(2 \pi f_0 \Omega B \epsilon_+\right)^2 \times \left[ \delta \left(  \omega \left(4 +30 \Delta \right) -\Omega \right) \delta(0) \right]~.
\label{trans_prob_lin_pol}
\end{eqnarray}
In a real experimental set up the observation time is finite. So we can regularize the Dirac delta function as $\delta(\omega)=\left[ \int_{-\frac{T}{2}}^{\frac{T}{2}} dt~~  e^{i \omega t} \right] = T$. Therefore, the transition rates become
\begin{eqnarray}
\lim\limits_{T \rightarrow \infty} \frac{1}{T}P_{0^\beta\rightarrow 2^\beta} &= &  \left(2 \pi f_0 \Omega A \epsilon_+\right)^2 \times \left[\delta\left( \omega \left(2 +9  \Delta \right)- \Omega \right)  \right]
\label{tm} \\
\lim\limits_{T \rightarrow \infty} \frac{1}{T}P_{0^\beta\rightarrow 4^\beta} &=&  \left(2 \pi f_0 \Omega B \epsilon_+\right)^2 \times \left[ \delta \left(  \omega \left(4 +30 \Delta \right) -\Omega \right)  \right]~.
\label{trlpgw}
\end{eqnarray}
\noindent Now we can analyze the above results. Firstly, the transition rates (\ref{tm}) and (\ref{trlpgw}) show that the detector resonantes with the GW at frequencies $\Omega=\omega \left(2 +9  \Delta \right) $ and $ \Omega=\omega \left(4 +30 \Delta \right)$ respectively due to the presence of the Dirac delta functions which make the transition rates non-zero only around those frequencies. Notice that the resonant frequencies for transitions from the ground state to the excited states get modified by the GUP parameter $\beta$.
Secondly, here we get two transitions instead of one. The transition from the ground state to the second excited state at $\Omega=2 \omega$ is already there in the standard HUP framework. Interestingly, the transitions from the ground state to the higher excited states (that is excited states higher than the second excited state) are only due to the presence of the GUP.
Further, from the expressions of $A$ and $B$, it is clear that terms both linear and quadratic in the dimensionless GUP parameter $\Delta$ will appear in the transition $P_{0^\beta\rightarrow 2^\beta}$. It is a good feature for detecting the presence of the GUP as linear dependence in $\Delta $ is easier to observe. The transition  $P_{0^\beta\rightarrow 4^\beta}$ shows that terms quadratic in $\Delta$ are important.  Also the expression for the transition amplitudes show that the transition probability $P_{0^\beta\rightarrow 2^\beta}$ has greater magnitude than $P_{0^\beta\rightarrow 4^\beta}$.

\noindent The above discussion indicates that these results can help to probe the presence of the GUP in the resonant detectors of GW. With the above results in place, we now make an estimate of the GUP parameter $\beta_0$. To do this, we note that the correction to the resonant frequency due to the GUP must be less than the resonant frequency itself. Hence, we have the inequality $9\Delta\omega < 2\omega$. This inequality can now be recast in the following form
\begin{equation}
\beta_0 < \frac{2}{9}\times\frac{M_p}{m}\times\frac{M_p c^2}{\hbar\omega}~.
\label{ineq}
\end{equation}
Substituting the values of the mass $m$ of the resonant bar detector which is of the order of one ton ($1.1\times10^3 ~kg$), the frequency $\omega$ of the detector which is of the order of 1kHz ($\omega/(2\pi)=900~Hz$), and the Planck mass $M_p c^2=1.2\times 10^{19}~GeV$, we get $\beta_0 < 1.4 \times 10^{28}$. This upper bound on the GUP parameter is stronger than the one obtained in \cite{fm} where data from the AURIGA gravitational bar detectors have been used to set limits to parameters of Planck-scale physics
which is $\beta_0 <3 \times 10^{33}$.
Interestingly, computing the correction to the resonant frequency $2\omega$ due to the GUP (which is $9\Delta\omega$) with the value of $\beta_0 = 10^{28}$, we find that $9\Delta\omega/(2\pi) \approx 1.3~kHz$. Hence, this simple calculation shows that the GUP modes can ring up in order to be detected by the resonant bar detectors.
However, if we set the more stringent bound $\beta_0 =10^{21}$ obtained in \cite{sdas},
we find that $9\Delta\omega/(2\pi) \approx 1.3\times10^{-7}~kHz$. These modes due to the GUP are indeed difficult to be detected by the resonant bar detectors having the sensitivity of the order of $10^{-19}$ \cite{piz}.



\subsection{Periodic circularly polarized GW}
Now we move on to another template of GW. The periodic GW signal with circular polarization can be conveniently expressed as
\begin{equation}
h_{jk} \left( t \right) = 2f_{0} \left[\varepsilon_{\times} \left( t \right) \sigma^1_{jk} + \varepsilon_{+}\left( t \right) \sigma^3_{jk}\right] 
\label{cir_pol}
\end{equation}
with $\varepsilon_{+} \left( t \right)  = \cos \Omega t $ and $\varepsilon_{\times} \left( t \right)  = \sin \Omega t $, $\Omega$ being the frequency of the GW. Here also  we shall impose the physical restriction on the frequency and the condition of finite observational time. Following the same mathematical procedure applied in case of the linearly polarized GWs, we can easily find out the transition rates for the GW template (\ref{cir_pol}) to be
\begin{eqnarray}
\lim\limits_{T \rightarrow \infty} \frac{1}{T}P_{0^\beta \rightarrow 2^\beta} &=& \left(2 \pi f_0 \Omega A\right)^2  \times \delta \left(\omega(2 +9 \Delta)- \Omega \right) \nonumber\\
\lim\limits_{T \rightarrow \infty} \frac{1}{T}P_{0^\beta \rightarrow 4^\beta} &=& \left(2\pi f_0 \Omega B \right)^2 \times  \delta \left( \omega (4+30 \Delta) -\Omega \right)~.
\label{11}
\end{eqnarray}
The above results show that the findings for the linearly polarized GWs also hold in case of the circularly polarized GW signals as well. Therefore, we can say that the circularly polarized GW signals are also good candidates to probe the presence of the GUP in the resonant GW detectors.

 
\subsection{Aperiodic linearly polarized GW: GW Burst}
The in-spiral neutron stars or black hole binaries generally gives rise to GWs with aperiodic signals. These astrophysical objects emit GW signals with huge amount of energy at the time of their merging or final ring-down. The duration of these signals are very small $10^{-3}$ sec$< \tau_g< 1$ sec. Such natural phenomena are commonly referred as bursts. In this discussion, we shall take approximated models of such violent and explosive astrophysical phenomena as follows
\begin{eqnarray}
h_{jk} \left(t\right) = 2f_{0} g \left( t \right) \left(\varepsilon_{\times}\sigma^1_{jk} + \varepsilon_{+}\sigma^3_{jk}\right).
\label{lin_pol_burst}
\end{eqnarray}
The above GW template contains both components of linear polarization. The smooth function $g \left( t \right)$  needs to go to zero rather fast for $|t| > \tau_{{\rm g}} $. Let us take a Gaussian form for the function $g(t)$
\begin{equation}
g \left(t\right) = e^{- t^{2}/ \tau_{g}^{2}}.
\label{burst_waveform_Gaussian}
\end{equation}
Note that $\tau_g \sim \frac{1}{f_{max}}$, where $f_{max}$ is the maximum value of a broad range of continuum spectrum of frequency. The GW burst contains a wide range of frequencies due to its small temporal duration \cite{sb2}.  
Now the Fourier decomposed modes of the GW burst can be written as 
\begin{eqnarray}
h_{jk} \left(t\right) = \frac{f_{0}}{\pi} \left(\varepsilon_{\times}\sigma^1_{jk} + \varepsilon_{+}\sigma^3_{jk}\right)  \int_{-\infty}^{+\infty} \tilde{g} \left( \Omega \right) e^{- i \Omega t}  d \Omega 
\label{lin_pol_burst_Gaussian}
\end{eqnarray}
where $\tilde{g} \left( \Omega \right) = \sqrt{\pi} \tau_{g} e^{- \left( \frac{\Omega \tau_{g}}{ 2} \right)^{2}}$ is the amplitude of the Fourier mode at frequency $\Omega$.
\noindent Now we can immediately find out the transition amplitudes using the template (\ref{lin_pol_burst_Gaussian}) in the general expression for the transition amplitudes (\ref{10}). This yields
\begin{eqnarray}
C_{0^\beta \rightarrow 2^\beta} &=& -2 i f_0 \varepsilon_{+} A \left(2\omega +9 \omega \Delta \right) \tilde{g}(2\omega +9 \omega \Delta) \nonumber\\ 
C_{0^\beta\rightarrow 4^\beta} &=&-2 i f_0 \varepsilon_{+} B \left(4\omega +30 \omega \Delta \right) \tilde{g}(4\omega +30 \omega \Delta).
\label{15}
\end{eqnarray}
The expression of $\tilde{g} \left( \Omega \right) $ leads to the following forms for the transition probabilities
\begin{eqnarray}
P_{0^\beta\rightarrow 2^\beta} &=& \left(2 \sqrt{\pi} f_0  \epsilon_+ A \tau_g \left(2\omega +9 \omega \Delta \right)\right)^2 e^{-2 \left\{\frac{2\omega +9 \omega \Delta}{2} \tau_g\right\}^2} \nonumber\\
P_{0^\beta\rightarrow 4^\beta} &=& \left(2 \sqrt{\pi} f_0  \epsilon_+ B \tau_g \left(4\omega +30 \omega \Delta \right)\right)^2 e^{-2 \left\{\frac{4\omega +30 \omega \Delta}{2} \tau_g\right\}^2} ~.
\label{12}
\end{eqnarray}
Before we end our discussion, we consider a more realistic GW waveform with modulated Gaussian function $g(t)$ as 
\begin{equation}
g \left(t\right) = e^{- t^{2}/ \tau_{g}^{2}}  \,  \sin \Omega_{0}t~.
\label{burst_waveform_sine_Gaussian}
\end{equation}
This represents a more realistic model of the GW burst signal. The Fourier transform of this function reads
\begin{eqnarray}
\tilde{g} \left( \Omega \right) = 2 \pi \int_{-\infty}^{+\infty} g(t) e^{i \Omega t} d \Omega= \frac{i \sqrt{\pi} \tau_{g}}{2} \left[ e^{- \left(\Omega - \Omega_{0}\right)^{2}\tau_{g}^{2}/4} - e^{- \left(\Omega + \Omega_{0}\right)^{2}\tau_{g}^{2}/4} \right].
\label{sine-Gaussian_Fourier}
\end{eqnarray}
Using this, we get the transition probabilities to be
\begin{eqnarray}
P_{0^\beta\rightarrow 2^\beta} &=& \left[ e^{- \left(2 \omega +9 \omega \Delta-\Omega_{0}\right)^{2}\tau_{g}^{2}/4} - e^{- \left(2 \omega +9\omega \Delta +\Omega_{0}\right)^{2}\tau_{g}^{2}/4} \right]^{2} \times  \left\{f_{0} \epsilon_+ A \sqrt{\pi}\tau_{g}\left(2 \omega+9\omega \Delta\right) \right\}^2
\label{gtdf} \nonumber\\
P_{0^\beta\rightarrow 4^\beta} &=& \left[ e^{- \left(4 \omega +30 \omega \Delta-\Omega_{0}\right)^{2}\tau_{g}^{2}/4} - e^{- \left(4 \omega +30\omega \Delta +\Omega_{0}\right)^{2}\tau_{g}^{2}/4} \right]^{2} \times  \left\{f_{0} \epsilon_+ B \sqrt{\pi}\tau_{g}\left(4 \omega+30\omega \Delta\right) \right\}^2~.
\label{gtdf2}
\end{eqnarray}
The above relations show that the two exponential terms in the transition amplitudes are almost equal and hence cancel each other in the sub-Hz bandpass region. But for the conditions $\frac{2 \omega +9 \omega \Delta-\Omega_{0}}{\omega}, \frac{4 \omega +30 \omega \Delta-\Omega_{0}}{\omega}<<1$, the second term will be negligible with respect to the first one. Hence the transition probabilities simplify to
\begin{eqnarray}
P_{0^\beta\rightarrow 2^\beta} \approx e^{- \left(2 \omega +9 \omega \Delta-\Omega_{0}\right)^{2}\tau_{g}^{2}/2} \times  \left\{f_{0} \epsilon_+ A \sqrt{\pi}\tau_{g}\left(2 \omega+9\omega \Delta\right) \right\}^2
\nonumber\\
P_{0^\beta\rightarrow 4^\beta} \approx e^{- \left(4 \omega +30 \omega \Delta-\Omega_{0}\right)^{2}\tau_{g}^{2}/2}  \times  \left\{f_{0} \epsilon_+ B \sqrt{\pi}\tau_{g}\left(4 \omega+30\omega \Delta\right) \right\}^2~.
\label{gtdf22}
\end{eqnarray}
Eq.(\ref{gtdf22}) is consistent with all the observations made for the periodic GW with linear polarizaton. 
The whole exercise reveal that there can be transitions between the states of the GW-HO system induced by the presence of the GUP correction in the Hamiltonian of the system. Such transitions do not take place in the HUP framework and owes it's origin to the GUP. Therefore, these results indicate a new window to probe the presence of quantum gravity effects. 


\section{Conclusion}
We now summarize our findings. Our analysis in this paper reveal that gravitational wave data from the resonant bar detectors may allow us to detect the existence of the generalized uncertainty principle.
The quantum mechanical description of the gravitational wave-harmonic oscillator system in presence of the generalized uncertainty principle shows noticeable changes in the resonant frequencies and transitions of the detectors. We now point out our findings. Firstly, the non-degenerate states of the one-dimensional harmonic oscillator get shifted with modified energy eigenvalues due to the presence of the generalized uncertainty principle. The perturbative treatment of the time independent Hamiltonian bears the signature of the generalized uncertainty principle. Then the incoming gravitational waves make transitions between the states of the generalized uncertainty principle modified states. Eventually one finds observable effects of the generalized uncertainty principle in the transition rates of the detector from the ground state to the excited states. From the exact forms of the transition rates we have made the following observations.
\begin{itemize}
	 
	\item The resonant frequencies $\Omega=\omega(2+9\Delta)$ and $\Omega=\omega(4+30\Delta)$ at which the resonant detector responds to the incoming gravitational waves get modified by the generalized uncertainty principle parameter $\beta$.
	We hope that these shifts in the resonant frequencies will be detectable in these detectors in the recent future if generalized uncertainty principle exists. This observation is quite similar with that of the noncommutative structure of space \cite{sg4}, \cite{sg5}.

	\item In the presence of generalized uncertainty principle, we find that there are more than one transitions possible from the ground state to the excited states. Incoming gravitational wave makes only one transition from the ground state to the second excited state in the standard Heisenberg uncertainty principle framework at $\Omega=2 \omega$. But in the framework of the generalized uncertainty principle, both the time independent as well as the time dependent part of the Hamiltonian bears it's signature. The time dependent part of the Hamiltonian allows transitions from the ground state to the higher excited states (higher than the second excited state).  
	
 	\item Both the linear as well as the quadratic terms in the dimensionless  generalized uncertainty principle parameter $\Delta$. The linear dependence in $\Delta$ is easier to detect. Though the transition $P_{0^\beta\rightarrow 4^\beta}$ contains terms quadratic in $\Delta$, it may serve to be a promising candidate to realize the existence of generalized uncertainty principle. It is also to be noted that $P_{0^\beta\rightarrow 2^\beta}$ has greater magnitude than $P_{0^\beta\rightarrow 4^\beta}$.

	\item Our analysis show that both linear and circularly polarized gravitational waves are the good candidates to probe the presence of the generalized uncertainty principle in the resonant detectors. This observation is valid for both the periodic and aperiodic signals as well.
	
	\end{itemize}

\noindent The observations made in this paper reveal that resonant detectors may allow in the near future to detect the existence of an underlying generalized uncertainty principle framework. Moreover in the recent literature \cite{sb}, a mathematical connection between the generalized uncertainty principle and the spatial noncommutative structure of space has been shown. Our analysis also indicates a similarity between the findings in these two frameworks.\\

\noindent Before we end our discussion, we would like to point out some perspectives for resonant bar detectors. So far no gravitational waves have been detected by these detectors because of their lower sensitivity than laser interferometry. The AURIGA detector at INFN, Italy is probably the only operational resonant bar detector. The most sensitive frequencies of resonant bar detectors is typically of the order of 1~kHz. Theoretical models estimate that events like collapsing and bouncing cores of supernovas can produce huge intensities of gravitational waves in the vicinity of 1-3~kHz \cite{ody}. Hence, these are the events which are likely to be observed by the resonant bar detectors. The present day strain sensitivity of these detectors is of the order of $3\times 10^{-19}$ \cite{piz}. This value can be used to roughly estimate the distance of an exploding supernova from Thorne's formula \cite{thor}, which gives a relation between the strain sensitivity ($h$), energy converted to gravitational waves
($\Delta{E}_{GW}$), characteristic frequency of the burst ($f$) and the distance ($d$) of the burst source from us. The formula reads
\begin{eqnarray}
h=2.7\times 10^{-20}\left[\frac{\Delta{E}_{GW}}{M_s c^2}\right]^{1/2}
\left[\frac{1~kHz}{f}\right]^{1/2} \left[\frac{10~Mpc}{d}\right]
\label{Thorne}
\end{eqnarray}
where $M_s$ is the solar mass. For supernova events, an optimistic value for the fraction of energy converted to GWs ($\Delta{E}_{GW}/(M_s c^2)$) would be around $7\times 10^{-4}$ \cite{stark}. Considering the sensitivity of resonant bar detectors to detect GW bursts to be around $h\sim 3\times 10^{-19}$ and the frequency of the burst to be around $900~Hz$, yields the value of $d$ to be 25~kpc. Such astrophysical events may not have taken place so far, else they would have been detected by the resonant bar detectors by now. Recent studies of the observed rate of supernova events at distances of $25~kpc$ is $0.114$ per year at $90~\%$ confidence limit \cite{fry, aga}, thereby indicating that the probability of detecting such events per year at current sensitivities would be very small.
To the best of our knowledge, attempts are being made to increase the sensitivity to $10^{-21}$ by using the full capability of these detectors to cool down to very low milli-Kelvin temperatures \cite{pize}. Further, new resonant detectors of different shape and much larger mass have also been thought of in order to increase their sensitivity; spherical resonant detectors are such candidates \cite{costa}.

\noindent In this paper, we have made an estimate of the GUP parameter $\beta_0$ which has led to an upper bound  $\beta_0<10^{28}$. This is a much stronger bound than that obtained in \cite{fm} which is $\beta_0<10^{33}$. Estimation of the frequency modes that can ring up due to the GUP with this value of the GUP parameter reveals that such modes should be detected once the resonant bar detectors detect gravitational waves. However,  with a value of the GUP parameter of the order of $10^{21}$ leads to frequency modes which would remain undetectable with the present bar detectors since they turn out to be smaller than the resonant frequency (without GUP) by seven orders of magnitude. The best bet to detect such modes would be in LIGO and LISA. 



\section*{Appendix}

Since the entire analysis in the paper revolves around eq.(\ref{e5}), we present a brief derivation of this result following \cite{Magg}.  
A GW bar detector can be idealized as a set of test masses to describe the interaction of the GW with it. 
Now the classical equation of motion of a test mass in a curved background described by the metric $g_{\mu \nu}$ (in the absence of any external non-gravitational force) is given by
 \begin{eqnarray}
\frac{d^2 x^{\mu}}{d \tau^2} + {\Gamma^{\mu}}_{\nu \rho} (x) \frac{dx^\nu}{d \tau} \frac{dx^\rho}{d \tau} = 0
 \label{geoeqn}
 \end{eqnarray}
where $\tau$ is the proper time parametrizing the curve $x^\mu(\tau)$. This is commonly known as the geodesic equation.

\noindent Now considering two nearby time-like geodesics, one parametrized by $x^\mu(\tau)$ and the other by $x^\mu(\tau)+\xi^\mu(\tau)$, the geodesic deviation equation reads
\begin{eqnarray}
\frac{D^2 \xi^\mu}{D\tau^2}=-{R^{\mu}}_{\nu \rho \sigma} \xi^\rho \frac{d x^\nu}{d \tau} \frac{d x ^\sigma}{d \tau}
\label{geodeveqn}
\end{eqnarray}
where the covariant derivative of a vector field $V^\mu(x)$ along the curve is given by $x^\mu(\tau)$
\begin{eqnarray}
\frac{D V^\mu}{D \tau}\equiv \frac{dV^\mu}{d\tau}+ {\Gamma^{\mu}}_{\nu \rho} V^\mu \frac{d x^\rho}{d \tau}~.
\label{covder}
\end{eqnarray}
\noindent Eq.(\ref{geodeveqn}) shows that two nearby time-like geodesics experience a tidal gravitational force determined by the Riemann tensor. 


\noindent Now an experimentalist in the laboratory normally uses the proper detector frame, where the test mass is free to move with respect to an origin. It turns out that using the geodesic deviation equation makes things simpler than using the geodesic equation in the proper detector frame (laboratory frame). Hence to describe the interaction of GWs with the detector, we use the geodesic deviation equation in the proper detector frame with a suitable gauge choice, namely, the transverse-traceless gauge, commonly known as the TT gauge.

\noindent It is to be now noted that, although GWs are relativistic, the detector (test particle) moves non-relativistically. Therefore, in the low velocity limit, it can be easily shown that eq.(\ref{geodeveqn}) in the proper detector frame yields up to linear order in metric perturbation $h$, the following equation
\begin{eqnarray}
\frac{d^2 \xi^i}{d\tau^2}= -c^2 {R^i}_{0j0}~\xi^j~.
\label{geodeveqnp}
\end{eqnarray}

\noindent We now need to compute the Riemann tensor $ {R^i}_{0j0}$ due to the GW in the proper detector frame. Using the fact that in the linearized theory, the Riemann tensor remains invariant rather than just covariant as in full general relativity, one can compute this in any frame. The TT frame where the GW have the simplest form, is the best choice to compute it. Hence, we have
\begin{eqnarray}
 R^i_{0j0}=-\frac{1}{2 c^2}\ddot h_{ij}^{TT}~.
\label{rt}
\end{eqnarray}
Therefore, the equation of the geodesic deviation in the proper detector frame takes the form 
\begin{eqnarray}
\ddot \xi^i= \frac{1}{2} \ddot h_{ij}^{TT} \xi^j~.
\label{gdepdf}
\end{eqnarray}
The above equation states that in the proper detector frame, the effects of the GW on a test particle of mass $m$ can be described in terms of a Newtonian force equation.

\noindent Before proceeding further we would like to point out two crucial features of 
eq.(\ref{gdepdf}).

\begin{itemize}
	 
\item The mathematical description of the response of the detector to GW can be framed in a purely Newtonian language. The Newtonian force due to GW is expressed in terms of $h_{ij}$ in the TT gauge.

\item In deriving the geodesic deviation equation, one needs to expand the Christoffel symbols to first order in $\xi$, neglecting all higher orders. This turns out to be valid only when $L \ll \lambda$, where $\lambda$ is the reduced wavelength of the GW and $L$ is the characteristic linear size of the detector. This condition is satisfied to a first approximation by resonant bar detectors.

\end{itemize}

\noindent Now  elastic theory of matter shows that the fundamental mode of a thin cylindrical bar is formally identical to a harmonic oscillator. Therefore, interaction of GW with the resonant bar detector can be recast as GW-HO system. Hence, the geodesic deviation equation in the proper detector frame for a two-dimensional HO of mass $m$ and intrinsic frequency $\varpi$ can be written as
\begin{eqnarray}
\ddot \xi^i + \varpi^2 \xi^i= \frac{1}{2} \ddot h_{ij}^{TT} \xi^j~;~ i,j=1,2~.
\label{hogde}
\end{eqnarray}
Now a resonant bar is a macroscopic object, weighting more than two tons. But the oscillations of the bar due to GWs are so small that a classical treatment is no longer adequate. The oscillations of the bar due to GW can be described in terms of the number of phonons that are excited. Therefore, to get a complete picture of the response of a resonant bar detector interacting with GW, we need to carry out a non-relativistic quantum mechanical analysis of a HO.


\section*{Acknowledgements} The authors would like to thank the referees for making critical comments which has improved the quality of the paper substantially.



\begin{thebibliography}{99}
\bibitem{ad} Amati D, Ciafaloni M and Veneziano G 1989 Phys. Lett. B 216 41.
\bibitem{kk} Konishi K, Paffuti G and Provero P 1990 Phys. Lett. B 234 276.
	
	
\bibitem{lqg1}	C. Rovelli, Living Rev. Relativity 1, 1 (1998).
\bibitem{lqg2}	S. Carlip, Rep. Prog. Phys. 64, 885 (2001).

\bibitem{gi} Girelli, F., Livine, E.R., Oriti, D. Nucl. Phys. B 708, 411 (2005).

\bibitem{bronstein}M.P.~Bronstein, 1936a. Kvantovanie gravitatsionnykh voln (Quantization of gravitational waves). Zh. Eksp. Teor. Fiz., 6, 195.
\bibitem{bronstein1}M.P.~Bronstein, 1936b. Quantentheorie schwacher Gravitationsfelder. Phys. Z. Sowjetunion, 9, 140–157.

\bibitem{mead} C.A. Mead, Physical Review 135 B (1964) 849-862.
\bibitem{mmg} M. Maggiore, Phys. Lett. B 319, (1993), 83-86.
\bibitem{fsc} F. Scardigli, Phys. Lett. B, 452, (1999), 39–44.
\bibitem{rjadler} R. J. Adler, D. I. Santiago,  Mod. Phys. Lett. A, 14, 20 (1999) 1371.
\bibitem{rd} R. J. Adler, P. Chen, and D. I. Santiago, Gen. Relativ. Gravit. 33, 2101 (2001).
\bibitem{rb} R. Banerjee and S. Ghosh, Phys. Lett. B 688, 224 (2010).
\bibitem{sg2} S. Gangopadhyay and A. Dutta, Gen. Relativ. Gravit. 46, 1661 (2014).
\bibitem{sg}S. Gangopadhyay, A. Dutta, and A. Saha, Gen. Rel. Grav. 46, 1661 (2014).
\bibitem{fs1} F. Scardigli, R. Casadio, Eur. Phys. J. C 75 (2015) 425.
\bibitem{sbbh} R. Mandal, S. Bhattacharyya, S. Gangopadhyay, 
Gen.Rel.Grav. 143 (2018) 50. 






 
 \bibitem{sdas} S.~Das and E. C.Vagenas, Phys.~Rev.~Lett. 101, 221301 (2008).
 \bibitem{sdas3} S. Das and E. C. Vagenas, Can. J. Phys. 87, 233 (2009).
\bibitem{sp} S. Das and S. Pramanik, Phys. Rev. D 86, 085004 (2012).
\bibitem{sb} S. Gangopadhyay, S. Bhattacharyya, Phys. Rev. D 99, 104010 (2019) .

\bibitem{kempf} A. Kempf, G. Mangano, R. B. Mann, Phys. Rev. D 52 (1995) 1108.
\bibitem{id} I. Dadic, L. Jonke, S. Meljanac, Phys. Rev. D, 67 (2003) 087701.
\bibitem{pb} P. Bosso, S. Das, R. B. Mann, Phys. Rev. D 96 (2017) 066008.

\bibitem{pb11} P.~Bosso, S.~Das, Annals of Phys. 396 (2018) 254-265.



\bibitem{bawaj} M.~Bawaj, et al., Nature Communications 6 (2015) 7503. 
\bibitem{zf} Z. W.~Feng, S.Z.~Yang, H.L.~Li, X.T.~Zu, Phys. Lett. B 768 (2017) 81-85.
\bibitem{bushev}P.A.~Bushev et al., Phys. Rev. D 100 (2019) 066020.
\bibitem{fs} F. Scardigli, Journal of Physics Conf. Series 1275 (2019), 012004.

\bibitem{vgnas}F.~Scardigli, G.~Lambiase, E.~Vagenas, Phys. Lett. B 767 (2017) 242.

\bibitem{pikov} I. Pikovski, M. R. Vanner, M. Aspelmeyer M. S. Kim and C. Brukner, Nature Phys. 8 (2012) 393-397. 
\bibitem{bosso1} P.~Bosso, S.~Das, I.~Pikovski, M.R.~Vanner, Phys. Rev. A 96 (2017) 023849. 
\bibitem{kumar} S.P.~Kumar, M.B.~Plenio, Phys. Rev. A 97 (2018) 063855.

\bibitem{bbh}B.P.~Abbott, et al., (LIGO Scientific Collaboration and Virgo Collaboration), Phys. Rev. Lett. 116 (2016) 061102.
\bibitem{bns}B.P.~Abbott, et al., (LIGO Scientific Collaboration and Virgo Collaboration), Phys. Rev. Lett. 119 (2017) 161101.
\bibitem{lig}J.~Aasi, et.al., (The LIGO Scientific Collaboration), Class.Quant.Grav. 32 (2015) 074001.
\bibitem{vir}F.~Acernese, et.al., Advanced Virgo: a second-generation interferometric gravitational wave detector, Class.Quant.Grav. 32 (2015) 024001.


	
\bibitem{jw1}J. Weber, Phys. Rev. Lett., 22 (24) (1969) 1320.
\bibitem{jw2}V. Ferrari, J. Weber et al Phys. Rev. D 25 (1982 ) 2471.
\bibitem{GW-detection_status}{\it Status of Gravitational Wave Detection},  Adalberto Giazotto,  in General Relativity and John Archibald Wheeler, Ciufolini and R.A. Matzner (eds.), Astrophysics and Space Science Library 367, Springer, 2010.	



\bibitem{sb2} A. Saha, S. Gangopadhyay, Phys. Lett. B 681 (2009) 96-99.
\bibitem{sg1}A. Saha, S. Gangopadhyay, S. Saha, Phys. Rev. D 83 (2011) 025004.
\bibitem{sg22}  A. Saha, S. Gangopadhyay, Class. Quant. Grav. 33 (2016) 205006.	
\bibitem{sg3} S. Gangopadhyay, A. Saha, S. Saha, Phys. Rev. D 97 (2018) 044015.
\bibitem{sg4} S. Bhattacharyya, S. Gangopadhyay, A. Saha, Class. Quantum Grav. 36 (2019) 055006.
\bibitem{sg5} S. Gangopadhyay, S. Bhattacharyya, A. Saha, Ukr. J. Phys. 2019. Vol. 64, No. 11.

\bibitem{fm} F. Marin, F. Marino, et al., Nature Phys. 9, 71 (2013).
\bibitem{fma} F. Marin, F. Marino, New Journal of Phys. 16, 085012 (2014).
\bibitem{rbm} P. Bosso, S. Das, R. B. Mann, Phys. Lett. B, 785 (2018).
 
\bibitem{kurt}K.~Gottfried, T.M.~Yan, ``Quantum Mechanics: Fundamentals", Springer-Verlag New York, 2003.
 
\bibitem{piz}G.~Pizzella, 2010 Search for Gravitational Waves with Resonant Detectors, In General Relativity and John Archibald Wheeler (Astrophys. Space Sci. Libr. vol 367)
ed I Ciufolini and R.A Matzner (Dordrecht: Springer) (doi:10.1007/978-90-481-3735-0-12).


\bibitem{ody}O.D.~Aguiar, Research in Astron. Astrophys. (2011) Vol.11, No. 1, 1-42.




 


\bibitem{thor}K.S.~Thorne, 1987, 300 years of Gravitation, S.W.~Hawking, W.~Israel,
(Eds.), Cambridge University Press, Cambridge, 330.

\bibitem{stark}Stark $\&$ Piran 1986, Proceedings of the 4th Marcel Grossmann Meeting on General Relativity, Elsevier, Amsterdam, page 327.




\bibitem{fry}C.L.~Fryer, K.C.B.~New, Living Rev. Relativity 14 (2011) 1.

\bibitem{aga}N.Y.~Agafonova, et.al., The Astrophysical Journal 802 (2015) 47. 




\bibitem{pize}G.~Pizzella, ``Status of Resonant Bar Detectors", Lecture given at the Conference on Gravitational Waves : A Challenge to Theoretical Astrophysics, Trieste (2000).


\bibitem{costa}C.F.~Da Silva Costa, O.D.~Aguiar, J.Phys.Conf.Ser. 484 (2014) 012012.

\bibitem{Magg}Michele Maggiore, {\it Gravitational Wave, Vol I, Theory and Experiments} Oxford University Press, 2008.

\end{thebibliography}
\end{document}